\documentclass[11pt]{article}
\usepackage{epsfig}
\def\arcm{\hbox{$^\prime$}}
\def\arcs{\hbox{$^{\prime\prime}$}}  
\def\etal{{\it et al.}}
\def\Om{$\Omega_{\rm matter}$}
\def\Omm{\Omega_{\rm matter}}
\def\AJ{{\em AJ}}
\def\ApJ{{\em ApJ}}
\def\ApJS{{\em ApJS}}
\def\gappr{\lower 3pt\hbox{$\buildrel > \over \sim\;$}} 

\begin{document}
\title{The Normal Cluster Weak Lensing Survey: Mass Profiles and M/L Ratios of Eight Clusters at z=0.2}

\author{David Wittman \\
Bell Laboratories, Lucent Technologies,
Murray Hill, NJ 07974 USA \\ \\
Ian Dell'Antonio \\
Physics Department, Brown University, Providence, RI 02912
USA \\ \\
Tony Tyson \\
Bell Laboratories, Lucent Technologies,
Murray Hill, NJ 07974 USA \\ \\
Gary Bernstein\\
Astronomy Department, University of Michigan, Ann Arbor, MI
48109 USA \\ \\
Philippe Fischer \\
University of Toronto, Dept. of Astronomy, Toronto, ON M5S 3H8
Canada \\ \\
Deano Smith \\
Science Department, Glenelg High School, Glenelg, MD 21737 USA}
\date{}

\maketitle
{\bf We present a survey of mass profiles and
mass-to-light ratios of eight typical galaxy clusters at a common
redshift ($z \sim 0.2$).  We use weak gravitational lensing as a probe
because it is unique in avoiding any assumptions about the dynamical
state of the clusters.  To avoid bias toward the rare and spectacular
clusters that are easy targets for lensing work, we selected an
ensemble of much more common clusters with moderate X-ray luminosity.
Although the survey is still in progress, two conclusions are
emerging: (1) within a cluster, mass follows light very closely on the
angular scales that we can measure, $0.2-2h^{-1}$ Mpc, and
(2) there is a significant cluster-to-cluster scatter in mass-to-light (M/L)
ratios despite uniformity of observing, reduction, and analysis
procedures.  We also derive an estimate of \Om\ based on extrapolation
from the mass properties of these typical clusters.  Finally, we
discuss the discovery of other clusters in our fields through their
lensing signal.  }

\section{Motivation and sample selection}
Everyone at this conference agrees that cluster masses are important
as cosmological probes and as clues to the formation process of
clusters and larger structures.  Yet most of the work on cluster
masses presented here has been based on the assumption that clusters
are in hydrostatic or virial equilibrium, while at the same time we
have seen many clusters that are clearly not in equilibrium.\cite{Sa}
Gravitational lensing can play an important role here because it is
one of the few methods that does not make dynamical
assumptions.\cite{Me}  However, most of the clusters examined
with lensing to date have been rare, spectacular clusters, and thus
are not representative of clusters in
general.

Therefore we are conducting a weak lensing survey of ``typical''
clusters, which we define here in terms of X-ray luminosity: $5 \times
10^{43} < L_X$ (0.3---3.0 keV) $< 7 \times 10^{44}$ $h^{-2}$ ergs
s$^{-1}$, or $0.2-3.0 L_X^*$.  We imposed three selection criteria
based on observing considerations: $ -60 < \delta < 0$ ({\it i.e.},
visible from Cerro Tololo Inter-American Observatory [CTIO]), $|b| >
20$, and no star brighter than $R=10$ lying within a 45\arcm\ diameter
field centered on the cluster (to minimize the area lost to bright
star haloes).  Finally, we imposed a redshift cut motivated by the
desire for as uniform a sample as possible, at $z \sim 0.2$, while
retaining $\sim 10$ clusters.  This resulted in the criterion $0.15 <
z < 0.30 $.  We applied these criteria to three catalogs: the Extended
Medium Sensitivity Survey\cite{Gi} (EMSS), the Einstein pointed
archive\cite{Ha}, and an early version of the Serendipitous
High-redshift Archival ROSAT Cluster (SHARC) survey\cite{Ro} (the two
SHARC clusters were not included in their final statistical cluster
sample, but were found in the course of the SHARC survey using the
methods cited).  The resulting eight clusters are listed in Table~1.

\begin{table}\label{tab-sample}
\begin{center}
\begin{tabular}{|l|l|c|c|c|c|cr|}
\hline
Name & Ref. & $\alpha$ (J2000) & $\delta$ & z & L$_X$ & l & b \\ \hline
MS 0419.0-3848 & 1 & 04:20:45 & -38:42 & 0.23 & 0.3 & 242 & -45 \\
Abell 3364     & 2 &05:47:34 & -31:53 & 0.16 & 6.5 & 237 & -27 \\
MS 0849.7-0521 & 1 & 08:52:17 & -05:33 & 0.19 & 1.3 & 233 & 24 \\
RX J100150.1-193550 & 3 & 10:01:50 & -19:36 & 0.23$^*$ & 1.3 & 258 & 28 \\
RX J111512-3807     & 3 & 11:15:12 & -38:07 & 0.19$^*$ & 0.5 & 283 & 21 \\
MS 1205.7-2921 & 1 & 12:08:20 & -29:38 & 0.17 & 1.0 & 292 & 32 \\
MS 1317.0-2111 & 1 & 13:19:44 & -21:27 & 0.16 & 1.5 & 312 & 41 \\
MS 2307.9-4328 & 1 & 23:10:46 & -43:12 & 0.29 & $<$1.0 & 348 & -64 \\\hline
\end{tabular}
\end{center}
\caption{The Normal Cluster Survey target list.  Redshifts marked with
an asterisk were determined by us with CTIO 4-m and 1.5-m spectroscopy as part
of the survey; all others were found in the literature. References:
(1) Gioia \etal\ 1990, (2) Harris \etal\ 1993 (3) Romer \etal\ 2000.  
X-ray luminosities are in units of $h^{-2} \times
10^{44}$ ergs~s$^{-1}$. For MS
2307, the listed $L_X$ is an upper limit because of the presence of a
foreground AGN within the X-ray error box. }
\end{table}

\section{Observations and data reduction}

We took deep multicolor images of each cluster at the CTIO's 4-m
Blanco telescope, with the Big Throughput Camera\cite{Wi98} mounted at
prime focus.  The BTC was designed to eliminate concerns about mass
sheet degeneracy with its large field.  The camera has a 2 $\times$ 2
array of back-illuminated 2048$\times$2048-pixel CCDs.  With pixels
subtending 0.43\arcs\ each and large gaps between the CCDs, the total
field of view comes to 35\arcm.  We covered the gaps and increased the
total field to 45\arcm\ with large dithers between exposures.  We
observed in the $B_jRI$ photometric system,\cite{Gu} with at least
$B_J$ and $R$ coverage on each cluster.

This paper is based on preliminary results from the eight clusters
listed.  We imaged two additional clusters, Abell 2744 and
MS0508.8-4523 (Abell 3322), which fit our criteria, but no results are
available for them yet.  These two clusters were also re-observed with
the Mosaic camera at the same telescope, and will be used to ensure
that no instrumental artifacts survive the data processing.  We also
observed most clusters in $I$ and some in Johnson-Cousins $V$ for the
sake of studying other cluster properties, but the current work
considers the $B_J$ and $R$ information only.  A forthcoming paper\cite{De}
presents the observations, lensing results,
and comparison with X-ray data and with simulations in great detail;
future papers will study the cluster luminosity functions and
alignment of cluster members.

The data reduction steps, especially those aspects related to optical
distortions and object shape measurement, are essentially the same as
that described elsewhere\cite{Wi00} in some detail for ``blank''
fields observed with the BTC.  The main difference is that here, we do
not attempt to combine shape measurements from different filters.
Since the $R$ images are deeper than the other filters, we simply use
the shapes from the $R$ images.  Because these fields are centered on
clusters, the analysis diverges from Ref. 9 beginning with the
background galaxy selection.

\section{Background galaxy selection and lensing analysis}

With clusters in the foreground, we make some effort to select
background galaxies based on color.  The color locus of cluster
ellipticals is easily identifiable in the color-magnitude diagram, so
we used only galaxies which were significantly bluer in an attempt to
eliminate cluster members from the lensing analysis.  A uniform
criterion $B_j - R < 1.5$ was used for all cluster fields.  We also imposed
a magnitude cut $22 < R < 26.5$ to eliminate foreground objects and
objects too faint to measure accurately.  We are currently working on
estimates of the residual contamination by cluster members.

Given this background galaxy sample, we use the model-independent
aperture densitometry method\cite{Fa} to estimate the
mass.  In discrete form, the crucial equation relating the galaxy
shapes to the mean surface
mass density $\overline{\Sigma}$ inside a given radius, minus the mean
surface density in a control annulus outside that radius is\cite{Fi}:

\begin{equation}
\overline \Sigma (r < \rm{r_0}) - \overline \Sigma (\rm{r_0} < r <
\rm{r_{max}})
= \Sigma_c  \sum^N_i \gamma_i ~ g(r_i) ~ {r_{\rm max} \over r_i}^2, ~~~~~
(\rm{r_0} < r_i < \rm{r_{max}})
\end{equation}
Here $\Sigma_c$ is the lensing critical density, $\gamma_i$ is the
tangential ellipticity of the $i$th background galaxy corrected for
seeing effects,\cite{Mi} $r_i$ is the projected distance from that
galaxy to the center of the mass distribution, g($r$) is a weighting
function, and N is the number of background galaxies in the annulus
$(\rm{r_0} < r < \rm{r_{max}})$.  We found the center of the mass
distribution from two-dimensional maps built up by repeatedly applying
a similar equation over a grid in x and y.  In most cases this
coincided with the brightest cluster galaxy and the center of the
X-ray emission, but in a few cases it was displaced by up to 1\arcm\
from the X-ray center.  $\Sigma_c$ depends on the source redshift
distribution and was calibrated for Abell 3364 by simulations with
full 3-dimensional ray tracing from model galaxies with a redshift and
morphology distribution consistent with the Hubble Deep Field and
recent Keck spectroscopy.  These simulations were in turn checked
against mass densitometry of clusters with known mass.

To obtain a comparable estimate of luminosity surface density, we also
measured the total light overdensity in $R$ band (with no color
selection of objects) in the same successive apertures, with respect to the
same control annulus.  $R$ band in the observed frame is very nearly
$V$ band in the rest frame, so we will quote M/L in rest-frame $V$.

\section{Mass and M/L ratios}

Cluster masses and M/L ratios within 1$h^{-1}$ Mpc are summarized in
Table~2.  The 1$h^{-1}$ Mpc radius was chosen to give reasonable
signal-to-noise even for the least massive clusters in the sample.
The separation between low M/L and high M/L clusters is roughly three
times the measurement error.  While a cluster-to-cluster scatter in
M/L has been seen in the literature,\cite{Me} it is not clear whether
the differences in the literature are entirely real or may be due in
part to the widely varied approaches to selection and analysis.  Here
an ensemble of clusters of similar X-ray luminosity, observed and
analyzed in a uniform way, shows similar scatter, strengthening the
case for real differences in M/L from cluster to cluster.

\begin{table}\label{tab-mass}
\begin{center}
\begin{tabular}{|l|c|c|}
\hline
Name & Mass ($h^{-1} \times 10^{14} M_\odot$) & M/L ($h M_\odot/L_\odot$) 
\\ \hline
MS 0419.0-3848 & 1.1 $\pm$ 0.4 & 160 $\pm$ 60 \\
Abell 3364     & 3.4 $\pm$ 0.4 & 230 $\pm$ 30 \\
MS 0849.7-0521 & 2.9 $\pm$ 0.5 & 350 $\pm$ 60 \\
RX J100150.1-193550 & 2.5 $\pm$ 0.4 & 340 $\pm$ 55\\
RX J111512-3807            & 1.5 $\pm$ 0.5 & 180 $\pm$ 60 \\
MS 1205.7-2921 & 2.0 $\pm$ 0.6 & 300 $\pm$ 90 \\
MS 1317.0-2111 & 1.2 $\pm$ 0.4 & 380 $\pm$ 130 \\
MS 2307.9-4328 & 2.3 $\pm$ 0.6 & 420 $\pm$ 110 \\
\hline
\end{tabular}
\end{center}
\caption{Cluster masses and M/L ratios within 1$h^{-1}$ Mpc.  M/L
refers to rest-frame $V$ band luminosity (observed in $R$ band).  Mass
errors are statistical only; there is an additional 10-20\%
calibration error affecting the entire ensemble.  M/L errors have been
propagated from the mass errors assuming small statistical errors on
the luminosity measurement.  M/L systematics (not quoted) include the
mass calibration error, plus smaller errors from background and
k-corrections.}
\end{table}

What might cause real differences in M/L?  It has been argued\cite{Ba}
that because the highest overdensities collapsed first, they now have
the oldest populations and therefore the highest M/L.  The highest
initial overdensities should now be the most massive clusters, so that
the clusters with higher M/L should also tend to be more massive.  We
face two problems in testing for this effect.  First, our clusters
were selected to lie in a narrow range of X-ray luminosity and
therefore presumably in a narrow mass range.  Second, the effect of
passive luminosity evolution is fairly small, while our errors in the
masses are about 20\%.  There is a hint that the more massive clusters
in Table~2 have higher M/L, but no conclusion can be reached given the
large error bars.  A sample including both massive and low-mass
clusters should be able to answer this question definitively.  We note
that our clusters do fall nicely on a curve of enclosed M/L versus
radius for all types of systems.\cite{Ba}

Figure 1 shows the mean mass and light profiles of Abell 3364.  The
other clusters have similar profiles but generally have lower
signal-to-noise.  Mass drops more steeply than isothermal at $r>1$ Mpc
(the apparent rise in the outermost point may be due to systematic
errors from the point-spread function at this very low level of
shear).  In these normal clusters, mass follows light surprisingly
well on all scales to which this technique is sensitive, out to at
least 2$h^{-1}$ Mpc; there is no evidence for an increase in M/L in
cluster outskirts.  This lends some legitimacy to extrapolating to
lower density environments since the mass and light for these normal
clusters are followed out to unprecedented small values: $\Sigma \sim 30$
M$_\odot$ pc$^{-2}$ mass density, and ${\mathcal L} \sim 0.01$ L$_\odot$
pc$^{-2}$, factors of 1000 and 10,000 smaller than customarily
encountered in the rich clusters studied in the past.

\begin{figure}\label{fig-mass}
\centerline{\psfig{file=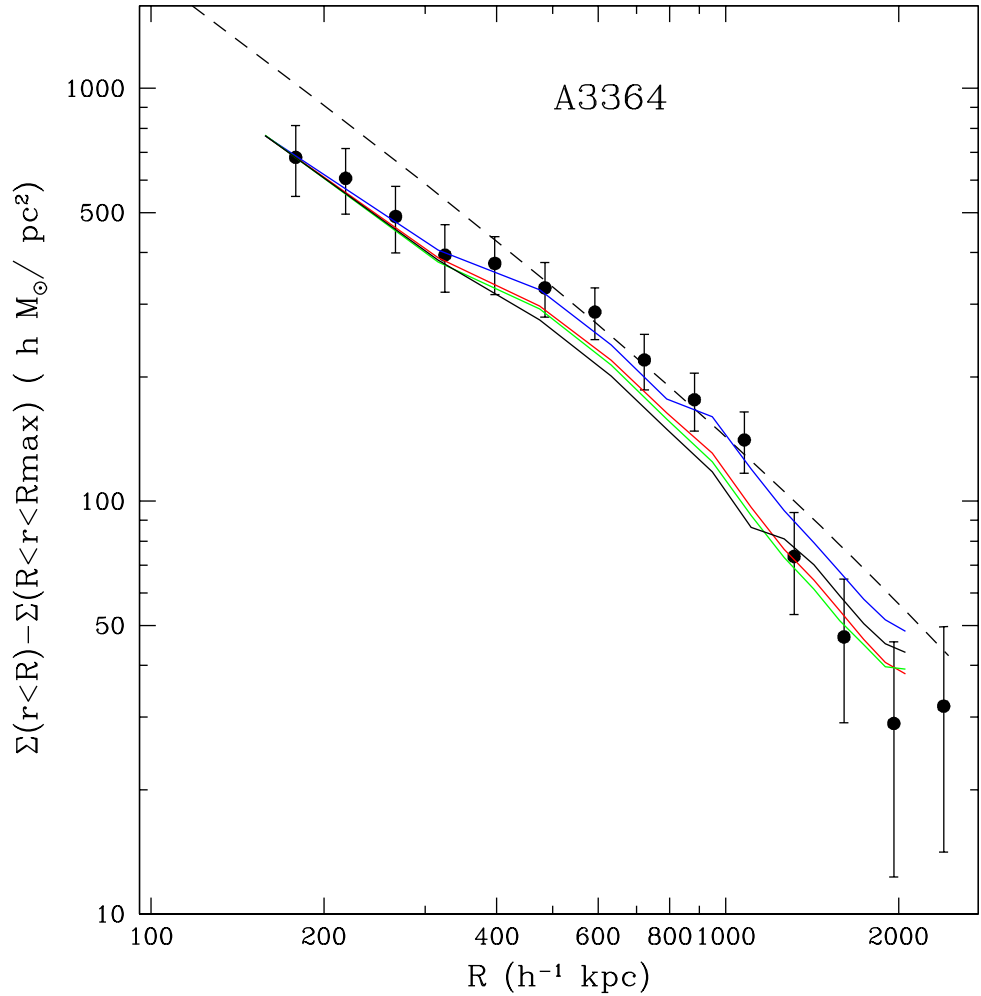,height=6in}}
\caption{Projected mass and light density profiles of Abell 3364.
Note that in the aperture densitometry method, error bars on adjacent
points are not independent.  The light profiles were observed in
observer-frame $B_j$ (blue line), $V$ (green), $R$ (red), and $I$
(black) filters, and computed in the same differential apertures
used for the mass.  The light profiles have each been shifted
vertically to intersect the innermost mass point, hence they are in
arbitrary units.  Mass follows light surprisingly well on all
measurable scales.  The dotted line shows the shape of an isothermal
profile, which is not quite a straight line with this estimator, to
guide the eye (it has not been fit to the data).  The two lowest mass
points are approaching the level of systematic error
estimated from the point-spread function. }
\end{figure}

\section{Estimate of \Om}

There are two ways to extrapolate from cluster mass information to
\Om, the mean mass density of the universe as a fraction of the
critical density.  Both methods assume that clusters are a fair sample
of the universe, so we must use some caution in interpreting the
results.  However, such an assumption is more plausible with this
sample of typical clusters than with previous samples which have been
based on a few very rich clusters.

One method can be applied immediately: {\it if} the cluster M/L holds for the
universe in general, we can simply scale the cluster M/L by the local
luminosity density, which is well known, and obtain the local mass
density.  In practice we scale the local luminosity density back to
z=0.2 using the recent luminosity evolution of galaxies.\cite{Ma}  Our
result is $\Omm = 0.19 \pm 0.03$, but we caution that the true
uncertainty lies in the M/L fair sample hypothesis rather than in the
statistical error given.  The previous section has shown that mass is
indeed well-correlated with light in the typical cluster environment,
but extremely low-density environments presumably emit less than their
share of light, so this might be considered a lower limit on \Om.

The second method will be applied in the near future: {\it if} the
cluster baryon fraction holds for the universe in general, we can
simply divide $\Omega_b$ by this fraction to obtain \Om.  $\Omega_b$,
the mean density of baryons in the universe, is well constrained by
Big Bang nucleosynthesis and so does not introduce great uncertainty
into the result.  However, this is generally considered an upper limit
to \Om\ because any census of baryons in a cluster is likely to miss
baryons (in currently undetectable forms such as black holes, planets,
etc.) rather than overestimate them.  We don't have any cluster baryon
fraction measurements as yet, but soon the combination of
Sunyaev-Zel'dovich effect measurements and lensing\cite{Ho} will
provide these.  For now we simply note that values in the literature
yield $\Omm \sim 0.3 h^{-2}$, which as an upper limit is a nice
complement to our value derived from luminosity scaling.

\section{Shear-selected clusters}

In the two-dimensional mass maps of the cluster fields, we found
significant additional mass clumps apart from the target clusters.
The density of these objects is roughly 4-7 deg$^{-2}$.  In many
cases, moderately bright galaxies can be seen clumped in these areas,
indicating the possible presence of a cluster or group.  In other
cases, there is no obvious optical counterpart.  We followed up
several of the cluster candidates with CTIO 4-m spectroscopy, and
confirmed that each is a cluster with a well-defined redshift, not a
chance projection.  In no case are they associated with the target
cluster, but they have the same general range in redshift, which is
not surprising considering that the background galaxy selection did
not vary.  These discoveries are an important step toward a
shear-selected sample of clusters and will be described in a
forthcoming paper.

\section{Summary}

Our weak gravitational lensing survey of ten clusters is still in
progress, but several conclusions are emerging.  We find a significant
cluster-to-cluster scatter in M/L, despite uniform observing,
reduction, and analysis procedures.  
Mass follows light very closely on scales from $\sim 200h^{-1}$ kpc
(our resolution limit) to at least 2$h^{-1}$ Mpc (the limit of our
field size), and to unprecedentedly low projected mass and light
surface densities.  This gives us some confidence in extrapolating to
even larger scales.  If the M/L of these clusters is representative of
the universe, we can scale by the local luminosity density to get
\Om\ \gappr 0.2.  In the near future, cluster baryon fractions will
be available from lensing plus Sunyaev-Zel'dovich effect
measurements\cite{Ho}, providing an estimate of \Om\ from baryon
scaling arguments.  The two scaling arguments really provide lower and
upper limits respectively; together they will bracket \Om\ if these
``normal'' clusters are a fair sample with respect to M/L and baryon
fraction.

The 45\arcm\ fields are large enough to include several several
serendipitous clusters, but more importantly we discovered them
{\it on the basis of shear alone}.  This demonstrates the feasibility
of a shear-selected sample of clusters, which would be immensely
valuable in determining the cluster mass function and avoiding the
biases toward emitted light which have always accompanied the study of
clusters.  Several surveys with this goal are already underway.

\section*{Acknowledgments}

We thank Alistair Walker, John Filhaber, David Rojas, Steve Heathcote,
and the staff at CTIO for their help in making the $BTC$ and 4-m
Blanco telescope system work in an optimal way and for their continued
support.  Thanks also to Bob Lee, Morley Blouke, and Pat Waddell for
their engineering help on the BTC project.  Initial development of the
BTC was partially funded by NSF SBIR grant AST86-17058 to Lassen
Research.


\end{document}